\documentclass[conference]{IEEEtran}
\IEEEoverridecommandlockouts
\usepackage{cite}
\usepackage{bbold}
\usepackage{amsmath,amssymb,amsfonts}
\usepackage{algpseudocode}
\usepackage{algorithm}
\usepackage{graphicx}
\usepackage{textcomp}
\usepackage{multirow}
\usepackage{dsfont}
\usepackage{eso-pic}
\usepackage{xcolor}
\usepackage{balance}
\usepackage{booktabs}
\usepackage{threeparttable}
\usepackage{todonotes}
\usepackage{wrapfig}
\usepackage{siunitx}
\def\BibTeX{{\rm B\kern-.05em{\sc i\kern-.025em b}\kern-.08em
    T\kern-.1667em\lower.7ex\hbox{E}\kern-.125emX}}
\usepackage{array}
\newcolumntype{P}[1]{>{\centering\arraybackslash}p{#1}}
    
\usepackage[hyperfootnotes=false]{hyperref} 

\newcommand\copyrighttext{%
  \footnotesize \textcopyright 2022 IEEE. Personal use of this material is permitted.
  Permission from IEEE must be obtained for all other uses, in any current or future
  media, including reprinting/republishing this material for advertising or promotional
  purposes, creating new collective works, for resale or redistribution to servers or
  lists, or reuse of any copyrighted component of this work in other works.
  DOI: \href{https://doi.org/10.1109/BigData55660.2022.10020860}{https://doi.org/10.1109/BigData55660.2022.10020860}}
\newcommand\copyrightnotice{%
\AddToShipoutPicture*{%
\put(45,30){%
\centering
\fbox{\parbox{\dimexpr\textwidth-\fboxsep-\fboxrule\relax}{\copyrighttext}}
}}}

\DeclareMathAlphabet{\altmathcal}{OMS}{cmsy}{m}{n}
    
\begin{document}

\title{Perona: Robust Infrastructure Fingerprinting for Resource-Efficient Big Data Analytics}

\author{
\IEEEauthorblockN{
Dominik Scheinert\IEEEauthorrefmark{1},
Soeren Becker\IEEEauthorrefmark{1},
Jonathan Bader\IEEEauthorrefmark{1},
Lauritz Thamsen\IEEEauthorrefmark{2},
Jonathan Will\IEEEauthorrefmark{1},
and Odej Kao\IEEEauthorrefmark{1}
}
\IEEEauthorblockA{
\IEEEauthorrefmark{1}
Technische Universit{\"a}t Berlin, Germany, \{firstname.lastname\}@tu-berlin.de}
\IEEEauthorblockA{
\IEEEauthorrefmark{2}
University of Glasgow, United Kingdom, lauritz.thamsen@glasgow.ac.uk}
}

\maketitle
\copyrightnotice

\begin{abstract}
Choosing a good resource configuration for big data analytics applications can be challenging, especially in cloud environments.
Automated approaches are desirable as poor decisions can reduce performance and raise costs.
The majority of existing automated approaches either build performance models from previous workload executions or conduct iterative resource configuration profiling until a near-optimal solution has been found.
In doing so, they only obtain an implicit understanding of the underlying infrastructure, which is difficult to transfer to alternative infrastructures and, thus, profiling and modeling insights are not sustained beyond very specific situations.

We present \emph{Perona}, a novel approach to robust infrastructure fingerprinting for usage in the context of big data analytics. 
Perona employs common sets and configurations of benchmarking tools for target resources, so that resulting benchmark metrics are directly comparable and ranking is enabled. 
Insignificant benchmark metrics are discarded by learning a low-dimensional representation of the input metric vector, and previous benchmark executions are taken into consideration for context-awareness as well, allowing to detect resource degradation.
We evaluate our approach both on data gathered from our own experiments as well as within related works for resource configuration optimization, demonstrating that Perona captures the characteristics from benchmark runs in a compact manner and produces representations that can be used directly.

\end{abstract}


\section{Introduction}
\label{sec:introduction}
Reliable, fast, and efficient data processing is crucial given the growing volumes of data in both industry and research.
These needs are often addressed by using distributed dataflow frameworks like Spark~\cite{Zaharia2010}, and Flink~\cite{Carbone2015}.
As these frameworks' handle parallelism, distribution, and fault tolerance, they make it easier for users to create scalable data-parallel programs.
The resulting applications can use a variety of compute clusters for data processing.

However, it is still difficult to choose and configure resources in a way that closely meets user-specific goals and constraints~\cite{RajanKCK16,cloudcomputingchallenges2018}.
Numerous strategies have been put forth to assist users, and they can be grouped into two categories:
Model-based techniques~\cite{MaoAMK16,RajanKCK16,ShahAKW19,AlSayehS19,KirchoffXMR19,ChenLLWZ21silhouette,ScheinertTZWAWK21,WillTSBK21,AlSayehMJPS22} often rely on access to historical performance data, however, historical workload execution data is not always available.
Search-based techniques~\cite{AlipourfardLCVY17,HsuNFM18,bilal2020finding,klimovic2018selecta,fekry2020accelerating,MendesCRG20,LiuXL20,SongZLSFDS21} conduct costly profiling runs prior to executing the actual workload utilizing all, or a fraction, of the input data to iteratively create performance models.

Often enough though, the optimized resource configuration is only relevant for the workload at hand. 
Information about the underlying infrastructure are solely obtained implicitly, i.e., by measuring the performance of the target workload in one execution context.
As a consequence, a thorough understanding of utilized resources and their capabilities is lacking and insights gained cannot be easily transferred to other contexts, for instance, when profiling new workloads with different resource demands. 
This requires repeated profiling overhead for reoccurring or comparable workloads that could be avoided, rendering current approaches less resource-efficient than they could be.

Addressing these limitations, we present \emph{Perona}, a novel approach to explicit and robust infrastructure fingerprinting. 
It motivates the usage of common sets and configurations of benchmarking tools to assess the full capabilities of target infrastructures and to make the obtained benchmarking metrics directly comparable.
This explicit fingerprinting operation transparently reveals the characteristics of resources and allows ranking them.
Perona discards irrelevant benchmarking metrics in a data-driven manner by learning a dense, low-dimensional representation of input metric vectors. 
With these, more sophisticated resource decisions can be made for big data analytics, e.g., with regard to scheduling or resource allocations.
To be able to assess a recent benchmark execution, our approach incorporates results of prior benchmark executions, which is particularly useful for detecting resource degradation. 

\emph{Contributions}. The contributions of this paper are:

\begin{itemize}
    \item A novel approach for incorporating infrastructure fingerprinting into model-based methods for optimized resource configuration of workloads through ranking of resources and detection of degrading resource behavior.
    \item A method for context-aware representation learning of benchmark metrics, thereby not only discarding insignificant features but also taking prior benchmark runs and corresponding machine metrics into account. 
    \item An openly available implementation\footnote{\url{https://github.com/dos-group/perona-infrastructure-fingerprinting}} of Perona which we evaluated with regard to common metrics and in interplay with resource configuration methods for distributed dataflows and scientific workflows. 
\end{itemize}

\emph{Outline}. \autoref{sec:related_work} discusses the related work.
\autoref{sec:approach} describes the three main aspects of our approach in detail. 
\autoref{sec:evaluation} presents the results of our evaluation.
\autoref{sec:conclusion} concludes the paper and gives an outlook on future work.
\section{Related Work}
\label{sec:related_work}
Various approaches have been proposed in the past for optimized configuration of cluster and cloud resources for big data analytics workloads, which can be classified into either 
\emph{Implicit Profiling-Based} or \emph{Explicit Benchmarking-Based} approaches, and will be presented in the following.

\subsection{Implicit Profiling-Based Approaches}

CherryPick~\cite{AlipourfardLCVY17} is an iterative search-based method capable of finding near-optimal cloud configurations for workloads, without the knowledge of previous executions.
It leverages Bayesian optimization to create performance predictions for different applications. 
Arrow~\cite{HsuNFM18} has been designed to improve CherryPick.
It also performs Bayesian optimization to find a near-optimal cloud configuration, but additionally incorporates low-level performance information to enrich the input data for the  optimization process.
Vanir~\cite{bilal2020finding} initially finds a suitable cloud configuration by analyzing performance metrics of benchmarking runs. 
It further utilizes a Mondrian-forest-based performance model and transfer learning techniques to identify better configurations.
Another approach is Selecta~\cite{klimovic2018selecta}, a tool that recommends near-optimal configurations by utilizing latent factor collaborative filtering on sparse metric data obtained from profiling runs.
This allows for predicting the performance of workloads on various configurations.
Paragon~\cite{delimitrou2013paragon} is a heterogeneity and interference-aware scheduler that uses different stages of profiling to identify the best machine for a given workload.
Initially, a few workloads are run offline on all server configurations and the performance results are normalized.
During runtime, a newly incoming workload is profiled for a minute on two of the target servers, followed by an estimation of its performance on the other servers.
Similarly, Quasar~\cite{delimitrou2014quasar} profiles incoming workload for a short period on a small number of servers.
The gathered profiling information is enriched with the information from the extensive offline profiling.
Unlike Paragon, Quasar integrates the rescaling impact into their estimations.

Summarizing, these approaches utilize profiling runs in order to optimize cloud and cluster configurations for a set of specific workloads, whereas with Perona we aim
to create a general infrastructure model that can be employed to optimize the resource configuration for arbitrary workloads.
\subsection{Explicit Benchmarking-Based Approaches}

Rupam~\cite{xu2018heterogeneity} is a heterogeneity-aware task scheduler for Apache Spark.
For hardware profiling, Rupam uses sysbench to measure CPU and I/O performance and Iperf for the network speed of a certain machine.
Further, it uses real-time resource utilization metrics to mitigate resource contention during execution.
For scheduling, Rupam uses a self-adaptable heuristic that considers the measured task resource usage and profiled machine attributes when assigning tasks to machines. 
With~\cite{zhang2013benchmarking}, a benchmark-based approach for performance modeling of MapReduce jobs is presented that separates the job tasks into six phases.
Then, a set of microbenchmarks is defined for representing the execution time of the phases, and executed on a test cluster to gather training data for the performance model.
This model only needs to be created once and can later be reused for different applications.

Further, in our recent work, we also used explicit benchmarking-based approaches to optimize resource management.
Tarema~\cite{BaderTarema21} is a system that dynamically allocates tasks to heterogeneous cluster nodes based on a ranking of resources.
Internally, it makes use of hardware profiling with a set of microbenchmarks, utilizes additional static hardware information, and computes group nodes with similar performance characteristics based on these profiled metrics.
Reshi~\cite{BaderReshi22} trains a recommender system that uses task resource traces and profiled infrastructure metrics as an input vector.
Several CPU, memory, and disk I/O benchmarks are executed on every node in the cluster, so that the recommender system is able to learn a model that can prioritize task-node pairs and execute the scheduling without knowing the actual task runtimes.
Lastly, Lotaru~\cite{BaderLotaru22} is a method that predicts task runtimes for heterogeneous clusters via locally obtained task profiles and without relying on the availability of any historical traces.
Again, Lotaru applies explicit benchmarking of the local and the target infrastructure to set an adjustment factor for predicting workload performance on the target infrastructure.

In contrast to Perona, the presented systems mainly apply an initial benchmarking phase, whereas we additionally compare and rank the learned representations of resource configurations while also enabling self-adaptive capabilities.
\begin{figure*}[ht]
    \centering
    \includegraphics[width=0.8\textwidth]{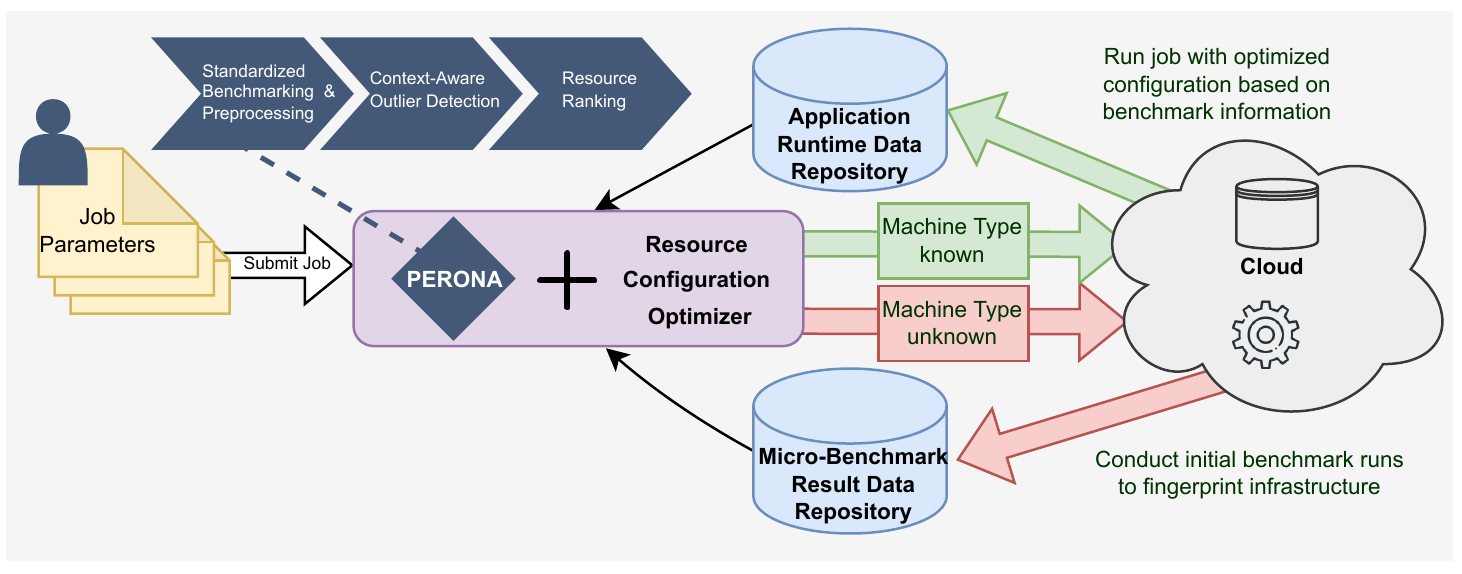}
    \caption{Overview of Perona. Designed as extension to common approaches for resource configuration optimization, target infrastructures are thoroughly benchmarked, relevant information are extracted, and anomalous benchmark executions due to resource degradation or failures are detected and reported.}
    \label{fig:overview}
\end{figure*}

\section{Approach}
\label{sec:approach}
This section presents the main ideas of our approach \emph{Perona} and how it can be used to explicitly fingerprint target infrastructures and extend existing resource configuration optimization solutions.
An overview is depicted in \autoref{fig:overview}, and the main variables used are summarized in~\autoref{tab:variabledefinitions}.

\begin{table}[hb]
\centering
\caption{Overview of main Variables}
    \begin{tabular}[t]{rl}
        \toprule
        \multicolumn{2}{c}{\emph{Problem Formalization}}\\
        \toprule
        $c_j$ & resource configuration $j$; $c_j \in C$\\
        $w_i$ & workload $i$; $w_i \in W$\\
        $y_{ij}$ & vector of performance measures for $w_i$ run with $c_j$\\
        \midrule
        \multicolumn{2}{c}{\emph{Approach}}\\
        \midrule
        $p_r$ & configuration template for resource aspect $r$; $p_r \in P$\\
        $m_k$ & machine type $k$; $m_k \in M$\\
        $b_r(t)$ & benchmark execution at time $t$ with $p_r$; $b_r(t) \in B(t)$\\
        $\vec{x}(t)$ & vector of all performance metrics in $B(t)$; $\vec{x}(t)\in \mathbb{R}^F$\\
        $\vec{x'}(t)$ & compact version of $\vec{x}(t)$ after preprocessing; $\vec{x'}(t)\in \mathbb{R}^{F'}$\\
        $\vec{c}(t)$ & dense learned encoding of $\vec{x'}(t)$; $\vec{c}(t)\in \mathbb{R}^{K}$\\
        $\vec{\underline{c}}(t)$ & neighborhood-inferred encoding; $\vec{\underline{c}}(t)\in \mathbb{R}^{K}$\\
        \bottomrule
    \end{tabular}
\label{tab:variabledefinitions}
\end{table}

\subsection{Overview}

Existing solutions to resource configuration optimization of big data analytics workloads commonly explore the resource configuration search space only indirectly, i.e., by means of running a target workload with different resource configurations and observing its performance.
While eventually an expedient approach, it does not provide an explicit and general understanding of resource configurations, which could be externalized and transferred for use in other contexts.
Hence, there lies potential in tackling this limitation.

We design a novel approach called Perona for explicit infrastructure fingerprinting, i.e., generalized benchmarking of resource configurations.
As illustrated in~\autoref{fig:overview}, it is at its core composed of three steps and associated components, which will be thoroughly described in the subsequent sections:
\begin{enumerate}
    \item A comprehensive inspection of target infrastructures through standardized sets and configurations of benchmarking tools, followed by an automated preprocessing of recorded benchmark metrics (\autoref{sec:approach_step1}).
    \item An internally employed graph-based modeling approach for representation learning of benchmark executions and context-aware outlier detection (\autoref{sec:approach_step2}).
    \item An opportunistic strategy for comparing and ranking learned representations of diverse resource configurations, which can be employed to automatically improve resource efficiency (\autoref{sec:approach_step3}).
\end{enumerate}
In the following, these ordered steps are explained in detail.

\begin{figure}[b]
    \centering
    \includegraphics[width=\columnwidth, keepaspectratio]{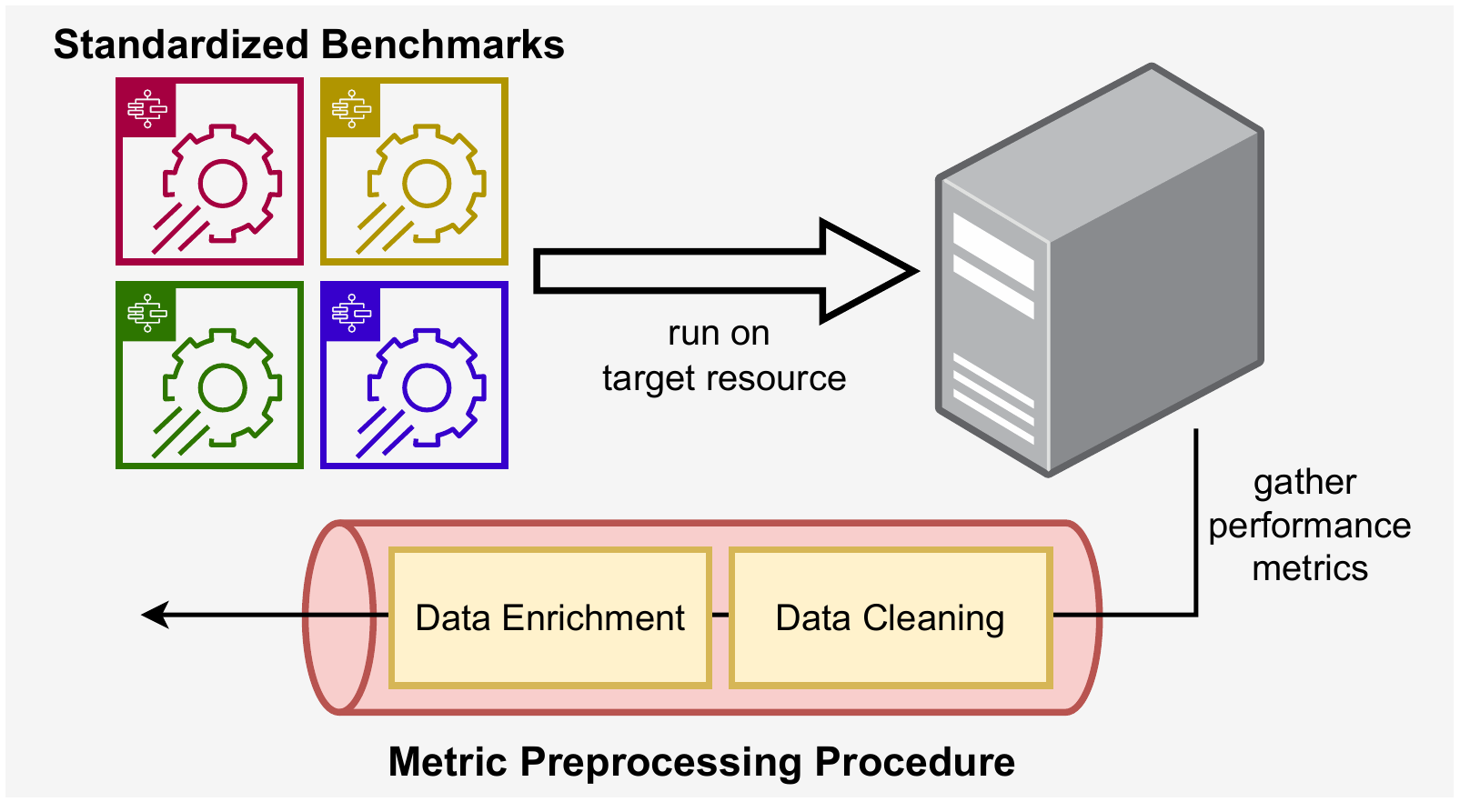}
    \caption{Standardized sets and configurations of benchmarking tools are used to holistically assess the characteristics of target machines and to ensure comparability across machines and benchmark executions.}
    \label{fig:approach_step1}
\end{figure}

\subsection{Standardized Resource Benchmarking}
\label{sec:approach_step1}

In order to sufficiently fingerprint a target infrastructure and carve out its individual characteristics, it is imperative to not only benchmark all relevant resources using dedicated tools, but also employ reasonable and fixed configurations for these benchmarking tools, so that individual runs are comparable across machine types and even infrastructures.
This idea is further depicted in~\autoref{fig:approach_step1}.
Let $r=1, \ldots, n$ index the up to $n$ different aspects of a resource configuration (e.g. memory, CPU, network) and $P=\{p_r\}_{r=1}^n$ be the associated set of configuration templates for benchmarking tools.
We then write $b^{(m_k)}_r(t)$ to denote the execution of a benchmarking tool at time $t$ with configuration template $p_r$ for resource aspect $r$ on machine $m_k \in M$, where the latter is part of the respective target infrastructure (e.g. the set of machines to profile, or a ready-to-use cluster of different machines).
The goal is then to initially fingerprint each so far unseen machine to obtain the complete set $B^{(m_k)}(t) = \{b^{(m_k)}_r(t)\}_{r=1}^n$, and latter on reschedule certain benchmarks if required. 
In this case, the marker $t$ is only approximate, since not all benchmark executions are necessarily conducted strictly in parallel.
With this, each machine is benchmarked the same way.
In the following and for the sake of simplicity, we will however refrain from using subscripts and superscripts if not strictly necessary.
This also makes sense given that our method is for the most part concerned with node-wise benchmark executions and benchmark-wise representation learning.

Each executed element of a set $B^(t)$ returns a variety of numerical performance metrics. 
We refer to the feature vector of all performance metrics associated with $B^(t)$ as $\vec{x}^(t)\in \mathbb{R}^F$, where $F$ denotes the total number of performance metrics and hence the feature dimension.
As of now, it is not yet clear which features are exactly relevant to sufficiently describe the benchmarked resource. 
Since a manual investigation can quickly become unmanageable, we are interested in automating this process.
Therefore, we apply a series of preprocessing steps to ease downstream modeling:
\begin{enumerate}
    \item \emph{Unification}: It can not be guaranteed that performance metrics are always issued in the same units. 
    Consequently, we ensure that all recordings of each individual performance metrics are among themselves comparable through unification of their associated units.
    \item \emph{Selection}: Some performance metrics might have less of a predictive value than others, which is why we only retain metrics with a standard deviation greater equal a configurable threshold.
    In addition, we require each metric to have at least two distinct historical values, otherwise its predictive character is questionable.
    \item \emph{Orientation}: Naturally, certain metrics (e.g. latency) are meant to be minimized whereas others are not (e.g. throughput).
    For our downstream modeling, we strive to equalize the various orientations as best as possible. 
    A performance metric shall be maximized if its maximum value is closer to its median than its minimum value, otherwise minimization is desired.
    Occasionally injecting synthetic stress into running benchmarks further helps in identifying the orientation of a metric.
\end{enumerate}
These steps help to reduce the dimensionality of feature vectors and to skip irrelevant individual features. 
Lastly, we enrich each feature vector by a one-hot encoding of the respective \emph{benchmark type} (tool + configuration). 
As a result, we obtain a compact feature vector $\vec{x'}(t)\in \mathbb{R}^{F'}$ with $F' \ll F$.

\textbf{Training Notes.} Since the aforementioned steps are designed to be stateful, the metrics associated with any $b_r(t) \in B(t)$ will be processed the same way and a feature vector of fixed size will be created.
In case of non-present features, e.g., an executed benchmark $b_r(t)$ evidently lacks the metrics generated by $b_s(t)$ with $r\neq s$, the missing values are filled with the so far observed average value of the metric of interest, which is a common machine learning practice. 

\subsection{End-to-End Contextual Representation Learning}
\label{sec:approach_step2}

\begin{figure}[b]
    \centering
    \includegraphics[width=\columnwidth, keepaspectratio]{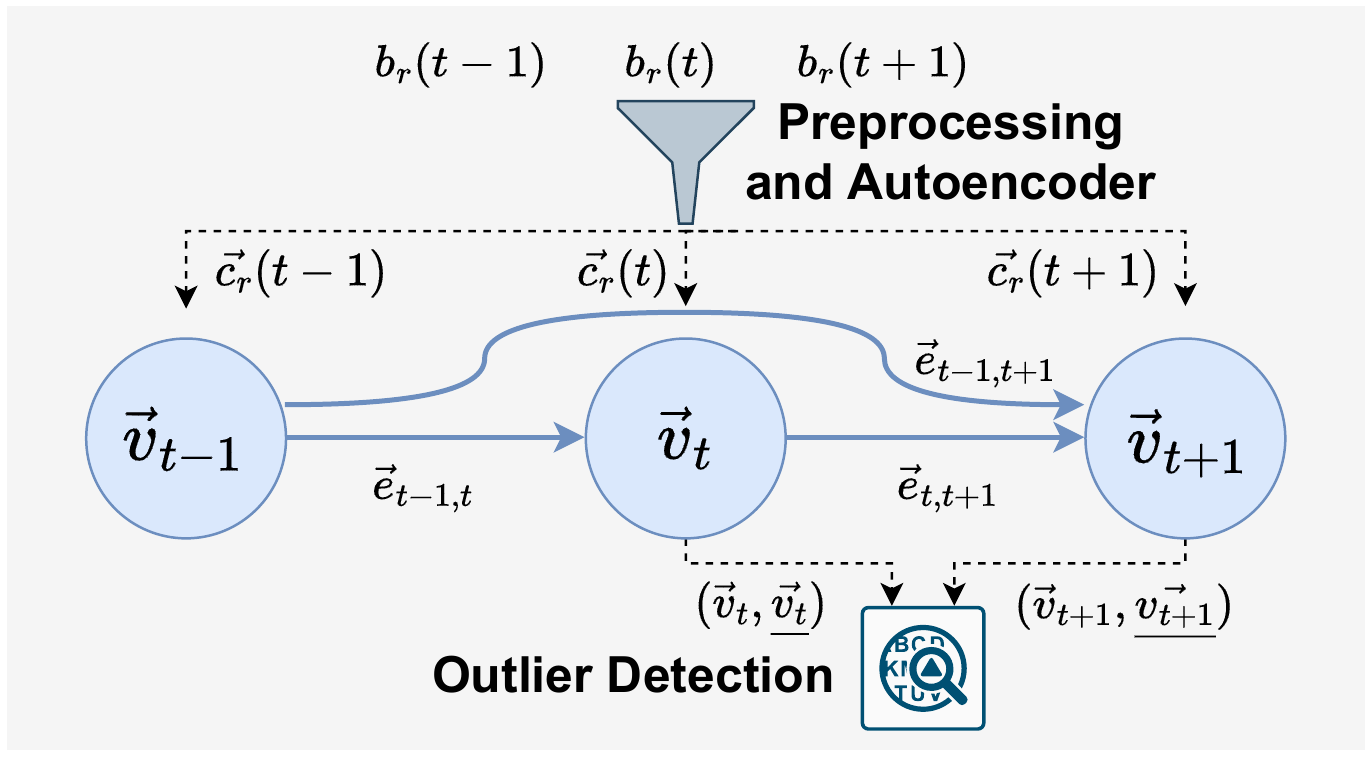}
    \caption{An autoencoder is employed for dimensionality reduction and feature extraction, such that key information are preserved. 
    The relationships of subsequent benchmark executions are exploited to detect anomalous behavior.}
    \label{fig:approach_step2}
\end{figure}

Correctly interpreting the results of a benchmark execution requires the consideration of relevant performance metrics and their contextualization with respect to prior benchmark executions. 
Hence, we propose a graph-based model that relies on dense representations of feature vectors learned by an encoder model and a decoder model.
It is illustrated in~\autoref{fig:approach_step2}.

Let $\vec{x'} \in \mathbb{R}^{F'}$ be a feature vector produced by our procedure detailed in~\autoref{sec:approach_step1} at an arbitrary point in time, a decoder network function $dec:\mathbb{R}^K\rightarrow \mathbb{R}^{F'}$ will try to reconstruct $\vec{x'}$ from the code $\vec{c}\in\mathbb{R}^K$ calculated by the encoder network function $enc:\mathbb{R}^{F'} \rightarrow \mathbb{R}^K$, such that $\min\Arrowvert \vec{x'} - dec(\vec{c}) \Arrowvert_p$ and $K \ll F'$ (in this context, $p$ denotes a desired $p$-norm).
This interaction enables the learning of meaningful and dense representations which can be used in downstream prediction tasks.
More precisely, the autoencoder learns the relevance of features, hence implementing an additional feature selection and realizing a dimensionality reduction.

At this point, we manage to grasp the relevant information from the original performance metric vector, yet we still lack a notion of normal and anomalous execution behavior since we focus on a single vector only. 
In a next step, we attempt to detect irregularities through consideration of previous benchmark executions.
Let $G=(V,E)$ be a directed and attributed graph that consists of a set of vertices $V=\{v_1, \ldots, v_n\}$ and a set of edges $E\subseteq \{(v_i,v_j)| v_i,v_j \in V\}$. 
An edge $e_{ij} \Leftrightarrow (v_i,v_j)\in E$ describes a directed connection between vertex $v_i$ and $v_j$. 
Thus, the node $v_j$ is then called a neighbor of node $v_i$, formally written as $j\in \altmathcal{N}(i)$.
Each node $v_i$ has an associated node feature vector $\vec{v_i}$, and each edge $e_{ij}$ can have an edge attribute vector $\vec{e_{ij}}$ as well. 
In the context of this work, $G$ is formed by establishing forward edge connections between chronologically sorted executions of the same benchmark type acting as nodes, with associated node feature vectors outputted by the encoder model $enc$. 
Furthermore, we use low-level metrics from the underlying compute instance obtained during a benchmark execution as well as various encodings of time intervals between each pair of benchmark executions to establish edge attributes. 
Note that graphs are composed per benchmark type and compute instance, in order to connect the relevant executions with each other.
The graph model $agg$ is then trained to predict the feature vector of a particular node via an aggregation of its respective neighborhood, i.e., through consideration of neighboring feature vectors and existing attributed edges:
\begin{equation*}
    \Vec{v_i}^{(k)} = \gamma^{(k)} \Big( \Vec{v_i}^{(k-1)}, \lambda_{j\in \altmathcal{N}(i)} \phi^{(k)} \Big( \Vec{v_i}^{(k-1)}, \Vec{v_j}^{(k-1)}, \vec{e_{ji}} \Big) \Big),
\end{equation*}
where $\lambda$ denotes a differentiable and permutation invariant function, $k$ denotes the number of hops, and both $\gamma$ and $\phi$ denote differentiable functions, e.g. feed-forward neural networks, which optionally alter the vector dimensionality.
When several such steps are performed, structural information is effectively used and passed through the graph.  
We temporarily denote the final aggregated version of a node feature vector $\vec{v_i}$ as $\vec{\underline{v_i}}$ -- the correctness of a benchmark execution can then be determined using a vector comparison.
We calculate the probability of a benchmark execution being anomalous as
\begin{equation*}
    \mathds{P}(\vec{v_i}) = \sigma (f_1(\vec{v_i} - \vec{\underline{v_i}})),
\end{equation*}
where $\sigma$ is the logistic function and $f_1$ is a non-linear transformation function with learnable parameters.

\textbf{Training Notes.} Due to the potentially significant imbalance of normal and anomalous benchmark executions, in our method implementation, we employ for the specific task of outlier detection a class-balanced focal loss~\cite{CuiJLSB19} (CBFL).
For the autoencoder, we minimize the reconstruction error, which is measured as the mean squared error (MSE) between input vectors and their reconstructed counterparts. 

\subsection{Aspect-Based Resource Ranking}
\label{sec:approach_step3}

\begin{figure}[b]
    \centering
    \includegraphics[width=\columnwidth, keepaspectratio]{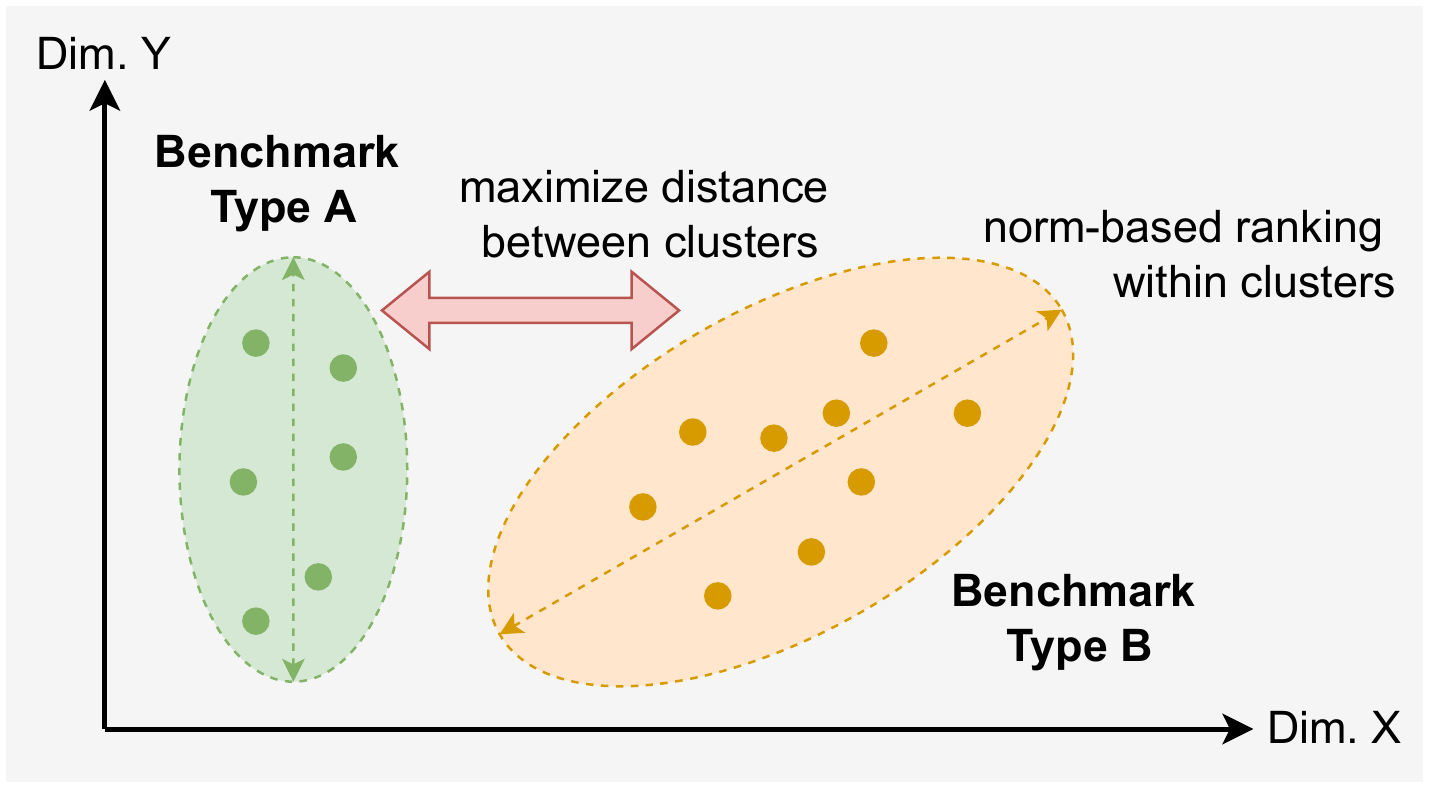}
    \caption{Our trainable function approximators are instructed to maximize the distance between benchmark clusters of learned encodings.
    Within each cluster, the encodings are positioned and ranked in terms of their vector norm.}
    \label{fig:approach_step3}
\end{figure}

In the previous step, we designed a way for learning latent benchmark execution encodings and detecting potential outliers. 
Yet, we so far have limited control over the general quality of learned encodings, which makes their direct usage questionable. 
In order to address this, we conceive several side tasks so that our model is instructed to learn more meaningful representations. 
More precisely, we demand the following:
\begin{itemize}
    \item \emph{Clustering}: Learned representations of one benchmark type shall be in close proximity in terms of cosine similarity, whereas the cosine distance of representations from unequal benchmark types shall be substantially increased.
    This leads to a class-based clustering of representations and is exemplarily depicted in~\autoref{fig:approach_step3}.
    \item \emph{Classification}: Closely related is the requirement that learned representations shall be sufficiently informative and hence usable for predicting the associated benchmark type after a simple linear transformation already.
    \item \emph{Ranking}: In order to embed our method into other approaches, it is imperative that learned representations are among each other comparable and rankable.
    We enforce this by deducing a pairwise ranking groundtruth based on a desired $p$-norm of preprocessed performance vectors, and demanding our learned representations to obey to the same norm-based ranking. 
    This is depicted within the clusters in~\autoref{fig:approach_step3} and consequently introduces a weak notion of order among representations.
\end{itemize}
The described additional learning tasks not only allow for more controlled representation learning, but also act as extra regularization component, since our model needs to optimize for multiple tasks simultaneously.
For a practical application of learned representations, e.g., in the context of scheduling problems where a suitable resource needs to be found for a target processing workload, one straightforward approach would then be to calculate a vector norm to obtain a score reflecting the quality of the resource in question.
This can be done for each resource aspect, allowing for a fine-granular assessment of resources and their capabilities.

\textbf{Training Notes.} For the clustering task, we employ a triplet margin loss~\cite{SchroffKP15} (TML) and combine it with a miner to evaluate especially challenging triplets.
The classification task is monitored using a cross entropy loss (CEL) operating on the output of the aforementioned linear transformation for obtaining class probabilities.
Finally, for the ranking of learned representations, we utilize a margin ranking loss (MRL) which evaluates the pairwise ranking of representations based on a previously deduced ranking groundtruth.
For pairs of normal representations, a ranking with minimal margins is sufficient, however, for unlike representations, we enforce a margin such that the anomalous representation is at least ranked smaller than the lowest normal representation observed so far.
\section{Evaluation}
\label{sec:evaluation}
This section presents our data acquisition procedure, our prototypical model implementation, and our experiments with accompanying discussion of the results.
All experiment-related artifacts are provided in our repository noted in~\autoref{sec:introduction}.

\subsection{Data Acquisition and Preparation}
All our experiments for data acquisition are conducted in K3s Kubernetes environments.  
In general, we use a fork of the Kubestone\footnote{\url{https://github.com/xridge/kubestone}, accessed: August 2022} benchmarking operator to assess the capabilities of target infrastructures with tools like \texttt{sysbench} (cpu and memory), \texttt{fio} (disk), \texttt{ioping} (disk), \texttt{qperf} (network), and \texttt{iperf3} (network).
These six benchmark types are used for our evaluation.
Upon successful execution, the benchmarking metrics are parsed from the associated results log via regex expressions and saved to a lightweight database.
To simulate varying benchmarking results, in some experiments, we occasionally impose stress on the respective hardware resource of interest using the ChaosMesh\footnote{\url{https://github.com/chaos-mesh/chaos-mesh}, accessed: August 2022} operator for chaos orchestration. 
For each series of experiments, we ensure that the Kubernetes nodes designated for benchmarking are drained as best as possible, i.e., all irrelevant Kubernetes Pods are moved to either the master node or a dedicated \emph{support} node. 
The latter also hosts the server Pods for client-server network benchmarks. 
Lastly, we enforce that at any point in time and across all target nodes, only one network benchmarking operation is executed in parallel. 
Over the total period of the experiment execution, we monitor and collect relevant resource metrics of target nodes using Prometheus and associated tools, and store them in our database as well. 
For automation of the aforementioned steps, we implemented a Kubernetes operator using the Kopf framework\footnote{\url{https://github.com/nolar/kopf}, accessed: August 2022} and work with Ansible scripts.  
All benchmark runs are quasi-randomly scheduled over the period of the total experiment time to possibly investigate different resource states.
Further technical details can be found in our repository.

\subsection{Model Implementation}
Each of our main functions $enc$, $dec$, and $agg$ is realized via soft computing, i.e., we rely on methods based on neural networks.
For the encoder $enc$ and decoder $dec$, we follow a straightforward design that we previously motivated~\cite{ScheinertTZWAWK21,scheinert2021enel}, although for the last layer of $dec$, we use a sigmoid transformation to account for our processed benchmark metrics.
For our function $agg$ used for graph-based message propagation, we average the outputs of two different graph transformations~\cite{ShiHFZWS21,tagconv} which are preceded by a configurable dropout on the adjacency matrix.
We subsequently utilize a non-linear activation~\cite{Klambauer2017}, configurable alpha-dropout~\cite{Klambauer2017}, and a final linear transformation and activation to alter the graph output.

The input to $enc$ is normalized to the range $(0, 1)$ feature-wise, where the boundaries are determined during training and used throughout inference.
The edge attributes used within $agg$ are prepared in the same manner.
In the extracted and to-be-processed graphs, each node has three predecessor nodes with which it is directly connected and thus can learn from.
During training, the model will attempt to optimize for multiple tasks simultaneously.
The various loss terms discussed in~\autoref{sec:approach} are hereby combined in an additive manner. 
For all norm-based operations, we utilize the $p$-norm with $p=10$.

In our experiments, we obtain a trained model after a hyperparameter search. 
The search space is depicted in~\autoref{tbl:clusterspecs_hyperopt}, and we sample 100 configurations from it using Ray Tune\footnote{\url{https://docs.ray.io/en/releases-1.13.0/tune/}, accessed: August 2022} with Optuna~\cite{Akiba2019}. 
More details can be found in the aforementioned repository. 
Note that although we run this on a dedicated machine equipped with a GPU (refer to~\autoref{tbl:clusterspecs_hyperopt}) to speed up the hyperoptimization procedure, the proposed model is generally compact and can also be trained on CPU-only machines, especially in light of periodic retrainings and identified near-optimal hyperparameters. 

\begin{table}
\centering
\caption{Resource Specifications \& Model Hyperoptimization}
\begin{tabular}[t]{cp{0.67\linewidth}}
    \toprule
    \multicolumn{2}{c}{\emph{Data Acquisition}}\\
    \midrule
    Software & K3s v1.21.10+k3s1, Kube-Prometheus-Stack 34.9\\
    & Kubestone v0.5.1, ChaosMesh 2.2.0, Ansible 2.12.7\\
    \midrule
    \multicolumn{2}{c}{\emph{Model Training}}\\
    \midrule
    CPU, vCores & Intel(R) Xeon(R) Silver 4208 CPU @ 2.10GHz, 8\\
    Memory & 45 GB RAM\\
    GPU & 1 x NVIDIA Quadro RTX 5000 (16 GB memory)\\
    Software & PyTorch 1.11.0, PyTorch Lightning 1.6.4\\ 
    & PyTorch Geometric 2.0.4, Ray Tune 1.13.0\\
    & Optuna 2.10.1, PyTorch Metric Learning 1.5.2\\
    \midrule
    \multicolumn{2}{c}{\emph{General Model Configuration and Hyperparameter Search Space}}\\
    \midrule
    Configuration & \#Epochs (max. 100), Hidden Dim. (32)\\
    & Batch Size (16), Optimizer (Adam)\\
    \#Attention-Heads & \{1, 3, 5\}\\
    Use beta & \{False, True\}\\
    Feature-Dropout & \{0\%, 10\%, 20\%\}\\
    Edge-Dropout & \{0\%, 10\%, 20\%\}\\
    Use root-weight & \{False, True\}\\
    CBFL $\gamma$ & \{0.5, 1.0, 2.0, 5.0\}\\
    CBFL $\beta$ & \{0.9, 0.99, 0.999, 0.9999\}\\
    Learning rate & \{$1e^{-1}, 1e^{-2}, 1e^{-3}$\}\\
    Weight decay & \{$1e^{-2}, 1e^{-3}, 1e^{-4}$\}\\
    \bottomrule
\end{tabular}
\label{tbl:clusterspecs_hyperopt}
\end{table}

\subsection{General Investigation of Fingerprinting Results}

For the data acquisition, we deploy a K3s Kubernetes cluster of five nodes in the Google Cloud Platform (GCP), where each node is a compute instance of machine type \emph{e2-medium} in the \emph{europe-west3} region.
Since one node is used as K3s master and another as support node, three nodes are used for benchmarking. 
Each of our six benchmarks is executed 100 times on each of the three nodes, where in 20\% of all cases, we impose stress on the respective Kubernetes benchmarking Pod. 
In total, we therefore conduct 1800 benchmark executions across nodes under varying conditions, hereby sufficiently capturing the variance in benchmarking results.
We then split the gathered data into 60\% training data, 20\% validation data, and 20\% test data in a stratified manner based on node names, benchmark types, and anomalous behavior. 
In the following, we present the results of our approach on the test data.

During the first step of our approach, 153 unique performance metrics (across all benchmark types) are preprocessed, enriched, and filtered down to 54 unique performance metrics.
The model then processes the input data and outputs learned representations as well as predicted class probabilities for outlier detection and benchmark type classification.
We observe a MSE on the test data of 0.01, demonstrating that the autoencoder can effectively recover the original input from a low-dimensional learned representation with high accuracy. 
Likewise, the learned representations are apparently sufficiently separated in the feature space, as we achieve with a simple linear transformation a 100\% accuracy on the benchmark type classification task. 
This confirms that learned representations indeed retain type-specific characteristics. 
Lastly, for the task of resource degradation detection and our opportunistic implementation of this specific module, we measure an F1 score of 0.93 for the normal class and an F1 score of 0.75 for the outlier class.
Moreover, the achieved weighted accuracy is 90\%.
While these results leave room for improvement, we conclude two things: First, the amount of data available for training has been fairly limited. Second, in practice, even false positive cases are not critical, since they can be used as trigger for subsequent benchmark runs, which then can either prove the original indication wrong or solidify it. 
In the latter case, one example of a reasonable action would be, for instance, the exclusion of the particular compromised node during workload scheduling.
With each benchmark type only running for a few seconds, the overhead is negligible, especially when compared to usually long-running actual workloads.

\subsection{Use Case: Iterative Optimization of Resource Configurations for Distributed Dataflows}
To assess the applicability of our learned representations, we resort to cloud configuration profiling methods for big data analytics and investigate how they benefit from our approach.
We consider \emph{CherryPick}~\cite{AlipourfardLCVY17}, which conducts profiling for a target workload based on Bayesian Optimization, as well as \emph{Arrow}~\cite{HsuNFM18}, a method that motivates Augmented Bayesian Optimization via consideration of low-level application metrics.
For both these methods, the overall objective is to find a near-optimal configuration in terms of execution costs while obeying to runtime constraints.
The idea now is to support this process by providing generalized infrastructure information.
Since for both methods we were not able to obtain the original source code, we re-implemented them to the best of our knowledge and provided them in our repository.
Furthermore, for Arrow, we simply replace their utilized low-level metrics with our computed scores from learned representations.
For both Arrow and CherryPick, we follow a simple integration: we weight the acquisition values produced by the respective acquisition function by a sum of products, where each product is composed of a factor that reflects the resource utilization of a particular configuration, and the other factor being the corresponding computed representation-based score.

\begin{figure}
    \centering
    \includegraphics[width=\columnwidth]{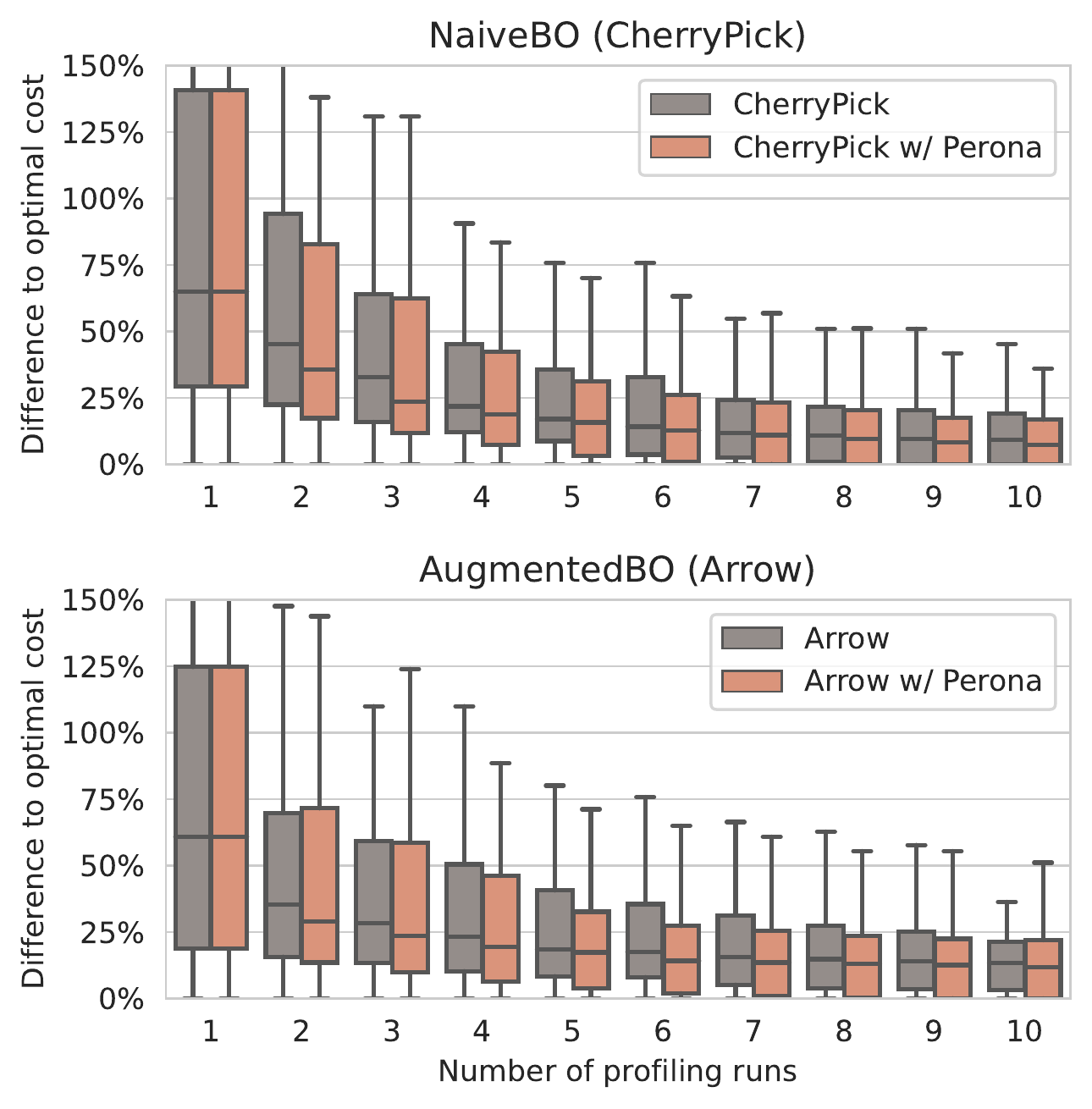}
    \caption{Illustration of the cheapest cloud configuration found under constraints, for both our baselines and their respective upgrades using Perona-based scores.}
    \label{fig:evaluation_df1}
\end{figure}

Both methods can be evaluated with a publicly available dataset\footnote{\url{https://github.com/oxhead/scout}, accessed: August 2022}, which includes data from 18 workloads running on 69 different configurations (scaleout, VM type) on Amazon Web Services (AWS) infrastructure in a multi-node environment (one run per configuration), resulting in a total of 1242 workload executions.
We derive the cost of each run using current prices\footnote{\url{https://calculator.aws/\#/createCalculator/EC2}, Accessed: August 2022} for AWS on-demand instances in the USA East Ohio region.
Beyond that, however, we still have to benchmark the machines used in this dataset via our own approach first.
We deploy a K3s Kubernetes cluster of 11 nodes in AWS, with two nodes (K3s master and support node) of machine type \emph{m4.large} and one each of size \emph{large}, \emph{xlarge}, and \emph{2xlarge} of machine types \emph{m4}, \emph{c4}, and \emph{r4}.
Each benchmark is executed 10 times on each node without stress injection, which totals to 540 benchmark executions used for learning representations.

In ~\autoref{fig:evaluation_df1} we present the cheapest valid cloud configurations found for subsequent profiling runs utilizing both baselines, respectively with and without the Perona extension. 
Here, the results of the first runs are identical, since at least one profiling run is required for Perona. 
As indicated by the results, an extension with Perona increases the median cost-effectiveness and hence improves cloud configuration profiling, especially for consecutive profiling runs where the model is enriched with further information. 
Although Perona tends to slightly increase the overall search cost and search time, eventually a more cost-effective configuration is found which is also less prone to timeouts induced by limited resources.

\subsection{Use Case: Adaptive Resource Management for Large-Scale Scientific Workflows}
In a last series of experiments, we investigate the applicability of our approach in conjunction with methods for scientific workflows.
More precisely, we examine if our learned representations can be employed instead of the manually selected benchmark metrics used within \emph{Tarema}\cite{BaderTarema21} and \emph{Lotaru}\cite{BaderLotaru22} while achieving similar performance.
Since the source code of both methods for scientific workloads is openly available, we can easily modify it and instruct the methods to consume our representations instead of their original benchmarking results.

As before, we first need to benchmark the machines used for evaluation in the respective works via our own approach.
We deploy a K3s Kubernetes cluster of five nodes in GCP, with two nodes (K3s master and support node) of the previously described machine type \emph{e2-medium} and one each of the \emph{n1-standard-4}, \emph{n2-standard-4}, and \emph{c2-standard-4} machine type.
Each benchmark is executed 10 times on each node without stress injection, which totals to 180 benchmark executions used for learning generalized representations.

For Lotaru, a system for runtime prediction of compute tasks that internally makes use of microbenchmarks, we substituted its microbenchmark values with our scores obtained from learned representations and adjusted the estimation process to fit for our used machines.
\autoref{tab:lotaru} presents the obtained results. 
The Naive approach, Online-M, and Online-P serve as baselines in the original paper and use a method that does not rely on benchmarking.
They are outperformed significantly by all approaches employing benchmarking.
Using Lotaru with Perona's fingerprinting leads to a median error increase of 1.74\%. 
However, the values reported for the 90th-percentile and 95th-percentile prediction error are similar to the results for Lotaru, indicating a satisfactory performance.
In the case of Tarema, a system that allocates cluster resources based on the tasks' resource usages and therefore uses microbenchmarks to group nodes with similar performance metrics, we mocked Tarema's group build with our previously obtained fingerprinting values.
As a result, Tarema created the same node groups, leading to the same overall workflow makespans.

\begin{table}
\centering
\caption{Fingerprinting results}
\begin{tabular}{c
    S[table-format = 2.4] 
    S[table-format = 2.4]
    S[table-format = 2.4] 
     S[table-format = 2.4]
      S[table-format = 2.4]
  }
\hline
       & \multicolumn{1}{|l|}{Naive} & \multicolumn{1}{l|}{Online-M} & \multicolumn{1}{l|}{Online-P} & \multicolumn{1}{l|}{Lotaru} & \multicolumn{1}{l|}{Perona} \\ \hline
Median & 0.6692425400491009         & 0.19547789073450178           & 0.19547789073450178           & 0.12368778482387402         & 0.14109173314021614         \\ \hline
P90    & 13.930279653897347         & 0.5966910636589577            & 0.5968593235157074            & 0.4741153530715184          & 0.4883234928027342         \\ \hline
P95    & 47.25881553469377          & 0.8010339573976514            & 0.808789857826177             & 0.793521193342131          & 0.7960764123370796          \\ \hline
\end{tabular}
\label{tab:lotaru}
\end{table}
\section{Conclusion}
\label{sec:conclusion}
This paper presents Perona, a novel approach to holistic infrastructure fingerprinting for improving resource efficiency of big data analytics.
Perona automates the assessment of target infrastructures via standardized sets and configurations of benchmarking tools and subsequent data preprocessing.
Our approach then utilizes representation learning and graph modeling for robust information extraction and detection of potential performance degradation. 
Learned representations can be directly used within existing resource configuration approaches, where they allow for similar or even superior optimization results.
Our evaluation on gathered data and in conjunction with existing methods shows that Perona successfully extracts and preserves relevant information, which in turn can be used in downstream optimization tasks. 

In the future, we plan to provide a unified framework for the management of data processing jobs by integrating Perona with recent resource configuration optimization methods.

\section*{Acknowledgments}
     \thanks{Funded by the Deutsche Forschungsgemeinschaft (DFG, German Research Foundation) as FONDA (Project 414984028, SFB 1404).}

\bibliographystyle{IEEEtran}
\bibliography{bib}

\end{document}